\documentclass{optica-article}

\journal{opticajournal} 

\articletype{Research Article}
\usepackage{siunitx}

\usepackage{acronym}

\acrodef{MBP}{Multi-beam ptychography}
\acrodef{SBP}{Single-beam ptychography}
\acrodef{FRC}{Fourier ring correlation}

\begin{document}

\title{Resolution-matched gain in Multi-beam ptychography}

\author{Runqing Yang,\authormark{1,*} Maik Kahnt,\authormark{2} and Pablo Villanueva-Perez\authormark{1}}

\address{\authormark{1}Division of Synchrotron Radiation Research and NanoLund, Department of Physics, Lund University, Lund, 22100, Sweden\\
\authormark{2}MAX IV Laboratory, Lund University, Box 118, 221 00, Lund, Sweden
}

\email{\authormark{*}runqing.yang@maxiv.lu.se} 


\begin{abstract*} 

In \ac{SBP}, reliable reconstruction requires overlap between adjacent scan positions. 
For a fixed probe size and overlap ratio, imaging a larger object area requires more measurements, limiting acquisition throughput.
\ac{MBP} illuminates multiple object regions simultaneously, potentially reducing the number of measurements. 
However, the achievable reduction at a given resolution remains unclear because additional data redundancy is needed to disentangle the contributions from multiple beams.
Here, we consider fully overlapped \ac{MBP}, in which the diffraction intensities from all beams fall on the same detector region and are summed incoherently.
Using numerical simulations, we adapted the scan step size to determine the minimum number of diffraction patterns required to achieve the same target resolution for different numbers of beams.
We compared \ac{SBP} with fully overlapped \ac{MBP} using up to 15 beams.
We define the gain as the measurement count for repeated \ac{SBP} scans divided by that for fully overlapped \ac{MBP}, at equal total imaged object area and identical achieved target resolution.
The required amount of raw measurements for \ac{MBP} increased approximately linearly with the number of beams used, but remained below that required for repeated \ac{SBP} scans over the same total imaged object area.
At a target resolution of 67\% of the spatial Nyquist frequency, evaluated using the one-bit threshold, fully overlapped \ac{MBP} achieved a gain of approximately two.
Detector cropping and lower resolution targets reduced this gain, indicating its dependence on the usable diffraction information and the reconstruction evaluation criterion.
Because the diffraction intensities from all beams overlap on the detector, the recorded patterns contain no spatial separation that can identify the individual beam contributions. 
The observed gain therefore provides a conservative benchmark, while separation within the detector plane offers a route toward larger gains.

\end{abstract*}

\section{Introduction}

Ptychography~\cite{rodenburg2008ptychography, pfeiffer2018x} is a widely used coherent imaging technique~\cite{miao2011coherent} that enables the high-resolution reconstruction of an object from far-field diffraction measurements.
In a typical experiment, a coherent probe scans across the object with overlap between adjacent illuminated positions, and a diffraction pattern is recorded at each position.
Recovering both the object and probe requires solving the phase problem, typically using iterative reconstruction algorithms~\cite{thibault2008high,maiden2009improved}. 
The key constraint for solving this non-convex and underconstrained problem is the redundancy introduced by scanning overlap between neighboring positions~\cite{moshtaghpour2025overlap,bunk2008influence}.
This redundancy allows neighboring measurements to constrain the same object region from different probe positions, thereby stabilizing phase retrieval.
Greater overlap can improve reconstruction quality, but requires more closely spaced scan positions and consequently longer overall acquisition times~\cite{hurst2010probe,bunk2008influence}.
While the imaged field of view is generally unlimited, the scanning nature of \ac{SBP} limits the throughput due to the time requirements.
This is most predominant when the use case requires large fields of view or the ability to resolve dynamic processes.

\ac{MBP} addresses this limitation by illuminating multiple spatially separated object regions simultaneously~\cite{batey2014information,bevis2016multiple,bevis2018multiple}, offering a route to increased imaging throughput~\cite{yao2020multi,wittwer2021upscaling}.
The probes are individually coherent but mutually incoherent, so their far-field diffraction intensities add incoherently in each measurement~\cite{aastrand2024adaptive,lyubomirskiy2022multi,li2024x,hirose2020multibeam,brooks2022temporal}.
Reconstruction must therefore recover multiple object regions and probes from the summed intensities.
Increasing the number of beams $N_{\mathrm{beam}}$ allows more object regions to be measured simultaneously, but also introduces more unknowns per recorded diffraction pattern.
The resulting trade-off between parallel acquisition and reconstruction redundancy~\cite{yao2020multi,hirose2020multibeam} determines how effectively \ac{MBP} can reduce the number of measurements.

Previous work has shown that the reconstruction quality decreases as the number of beams increases and that the throughput gain eventually saturates~\cite{penagos2025multiplexing}. 
These findings motivate quantifying the measurement reduction achievable when reconstruction resolution is held fixed.
Here, we focus on fully overlapped \ac{MBP}, in which the diffraction intensities from all beams are centered on the same detector position (Fig.~\ref{fig:setup}(a)).
This geometry provides no detector-plane spatial separation to distinguish the individual beam contributions, making the reconstruction particularly challenging.
Studying this geometry, therefore, tests whether \ac{MBP} can reduce the number of diffraction patterns required to image the same object area at the same resolution despite this additional reconstruction challenge.

To quantify how the required redundancy within the recorded data changes with the number of beams, we first formulate the forward model for fully overlapped \ac{MBP}. 
At the \(m\)th scan point, \(\mathbf{R}_m\) denotes the scan position and \(\mathbf{s}_n\) denotes the relative offset of the \(n\)th probe. 
For mutually incoherent beams, the measured far-field intensity is
\begin{equation}
I_m(\mathbf{q})
=
\sum_{n=1}^{N_{\mathrm{beam}}}
\left|
\mathcal{F}_{\mathbf{x}\rightarrow\mathbf{q}}
\left[
P_n(\mathbf{x}-\mathbf{R}_m-\mathbf{s}_n)O(\mathbf{x})
\right]
\right|^2 ,
\label{eq:mbp_forward}
\end{equation}
where \(\mathbf{x}\) is the object-plane coordinate, \(\mathbf{q}\) is the reciprocal-space coordinate, \(P_n\) is the complex probe function of the \(n\)th beam, \(O\) is the object transmission function, and \(N_{\mathrm{beam}}\) is the number of beams.
To relate this forward model to the required scan redundancy, we use the information oversampling ratio proposed for \ac{MBP}~\cite{penagos2025multiplexing}, which compares the number of measured intensity values with the number of unknown object and probe variables.
The scaling can therefore be written as
\begin{equation}
\sigma_{\mathrm{MBP}}
\propto
\frac{ MN_{\mathrm{pix}}}
{N_{\mathrm{beam}} N_{\mathrm{unknown,loc}}},
\label{eq:info_ratio}
\end{equation}
where \(M\) is the number of recorded diffraction patterns, \(N_{\mathrm{pix}}\) is the total number of pixels in one recorded diffraction pattern, and \(N_{\mathrm{unknown,loc}}\) denotes the number of unknown object and probe associated associated with one beam.
Increasing \(N_{\mathrm{beam}}\) therefore requires additional scan redundancy to compensate for the reduction in the information oversampling ratio.
Insufficient redundancy can degrade the reconstruction resolution or even prevent convergence.
Therefore, we investigate how much additional redundancy is required as the number of beams increases to maintain a given reconstruction resolution.

\section{Method}

A resolution-matched comparison requires consistent simulation conditions and a common reconstruction criterion.
This section describes these conditions and the reconstruction procedure used to determine the required number of diffraction patterns.

\subsection{Simulation model}

\begin{figure}[ht!]
\centering\includegraphics[width=\linewidth]{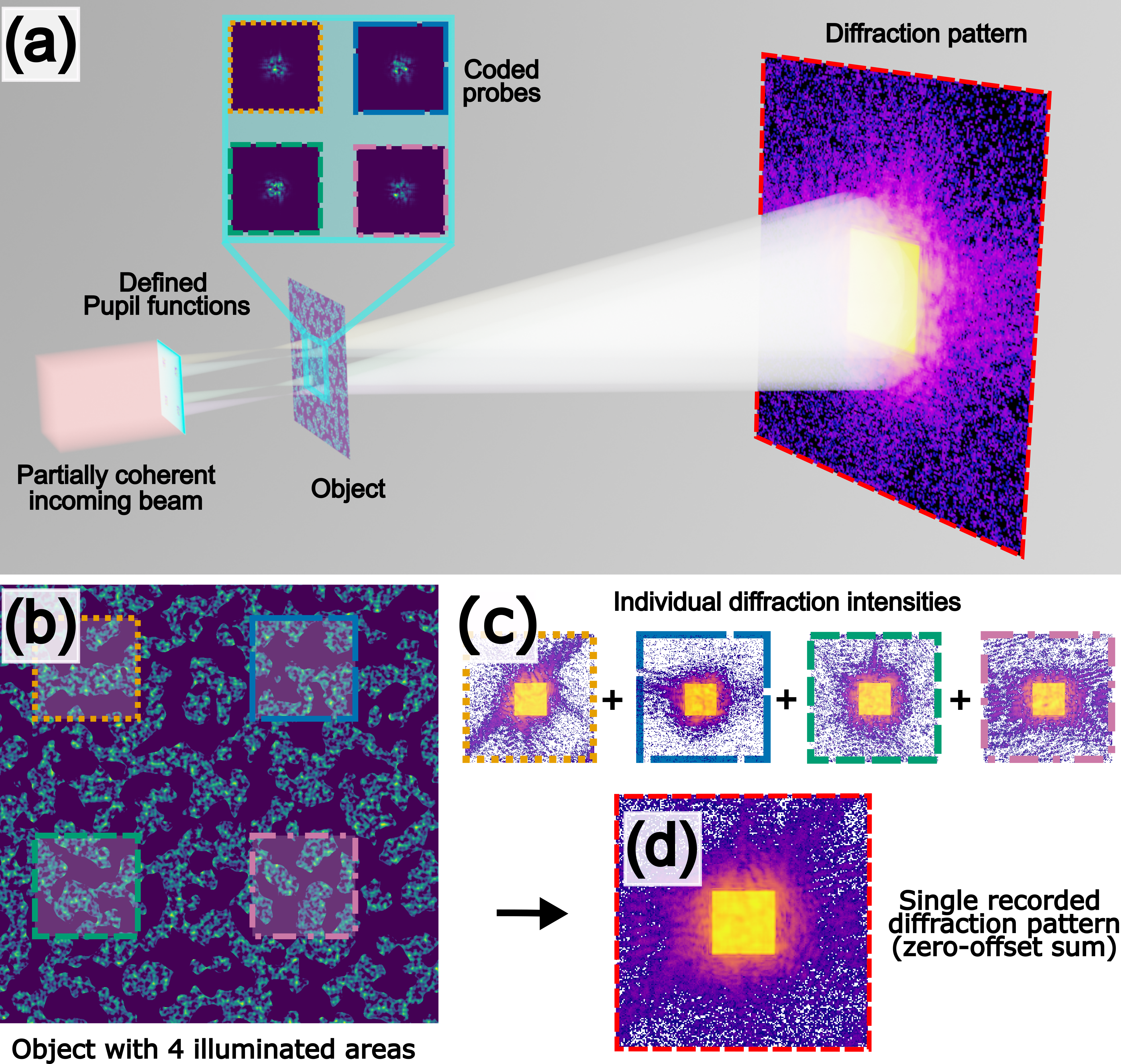}
\caption{Simulation geometry for fully overlapped \ac{MBP}. 
(a) Schematic of the forward model. 
(b) Object-plane view showing 4 illuminated regions in one multi-beam exposure. 
(c) Individual diffraction intensities generated by the corresponding exit waves, shown on a logarithmic scale. 
(d) Single recorded diffraction pattern obtained by the zero-offset incoherent sum of the individual diffraction intensities.}
\label{fig:setup}
\end{figure}

Fig.~\ref{fig:setup} illustrates the fully overlapped \ac{MBP} geometry used in the simulations.
We generated a statistically uniform complex-valued Perlin noise object~\cite{Abdolhoseini2019Neuron} and structured probes with Hadamard-like~\cite{hadamard1893resolution} phase modulations, following previously described procedures~\cite{yang2026optimizing}.
Each probe illuminated a separate $512\times512$-pixel object region.
Neighboring regions were separated by 522 pixels center-to-center, leaving a 10-pixel gap between them.
The same probe and region pairings were retained across comparisons.
At the object plane, each probe had an effective radius of 61 pixels and contained $10^7$ photons per exposure.
The total incident photon number per scan position was therefore $N_{\mathrm{beam}}\times10^7$.
At each scan position, we calculated the far-field diffraction intensity from each exit wave and applied Poisson noise independently.
The resulting intensities were then summed, producing the fully overlapped diffraction patterns described by Eq.~\ref{eq:mbp_forward}.

\subsection{Oversampling factor study}

The detector oversampling factor determines how finely the diffraction signal is sampled on the detector and can affect reconstruction quality~\cite{miao1997image,miao2000oversampling}.
A suitable sampling condition is therefore needed to compare different \ac{MBP} configurations consistently.
Insufficient oversampling can degrade reconstruction quality, resulting in a requirement for finer scan steps to compensate by providing additional redundancy.
Increasing the oversampling factor generally improves sampling of the diffraction pattern, but also increases the required detector distance and computational cost~\cite{chushkin2013upsampling}.
We therefore compared reconstruction quality at different oversampling factors to select a common sampling condition for the comparisons.

The detector oversampling factor was defined as
\begin{equation}
S=\frac{\lambda z}{Dp},
\label{eq:oversampling_factor}
\end{equation}
where \(\lambda\) is the wavelength, \(z\) is the object-to-detector distance, \(D\) is the effective probe diameter, and \(p\) is the detector pixel size. The oversampling factor was varied by changing \(z\), while \(\lambda\), \(D\), and \(p\) remained fixed. The corresponding real-space pixel size was 
\begin{equation}
\Delta x=\frac{\lambda z}{N_{\mathrm{det}}p},
\label{eq:real_space_sampling}
\end{equation}
where \(N_{\mathrm{det}}\) is the linear size of the recorded diffraction pattern in pixels.

To keep \(\Delta x\) constant across the oversampling factor study, \(N_{\mathrm{det}}\) was adjusted in proportion to \(z\). 
The probe array was zero-padded to the resulting diffraction pattern size without changing its effective diameter.
The object array was padded correspondingly to maintain consistent dimensions in the forward model. 
For each beam configuration, the same scan positions, the same total number of diffraction patterns and the same number of photons in the illuminating probe were used throughout the oversampling factor study.

\begin{figure}[!htbp]
\centering
\centering\includegraphics[width=\linewidth]{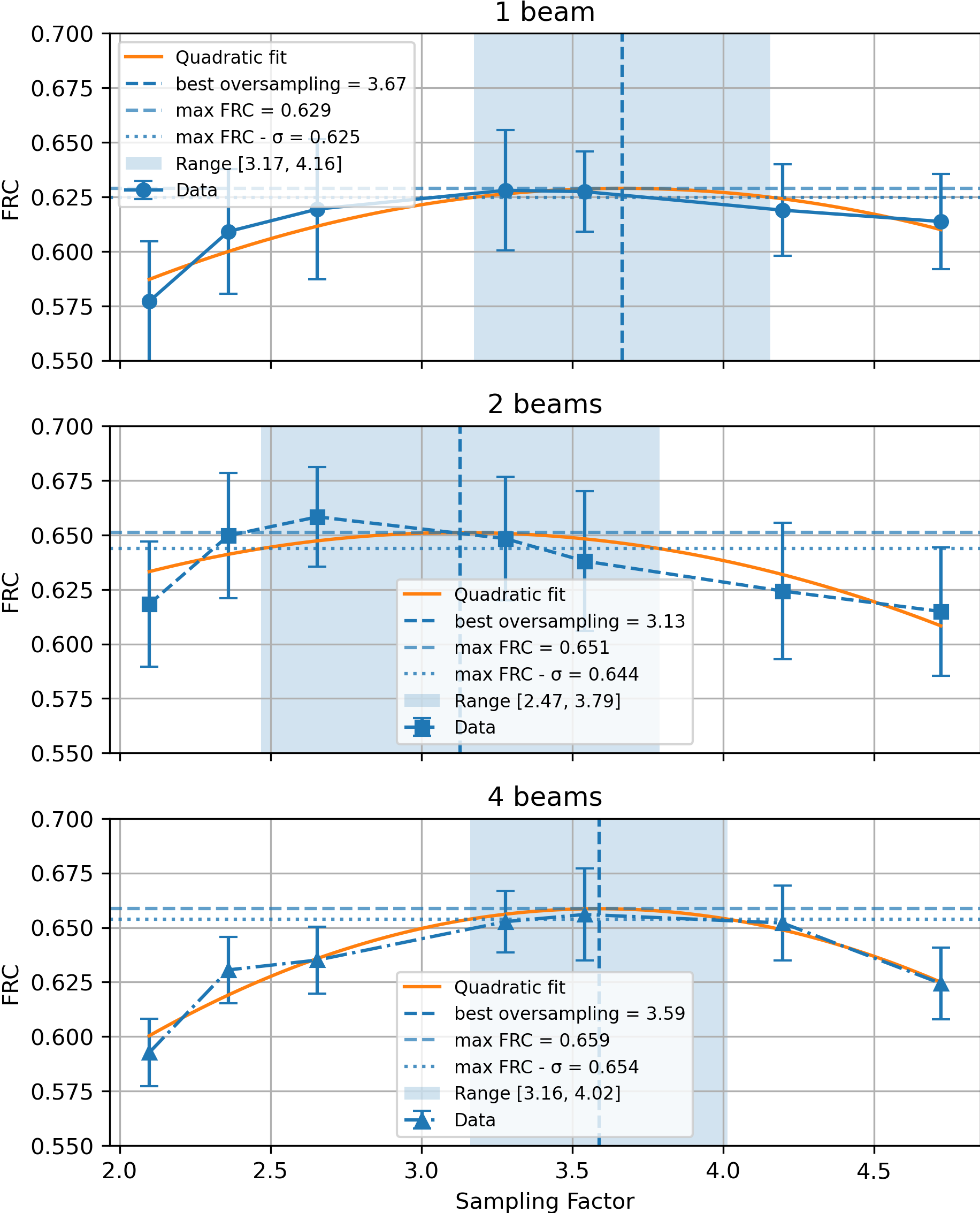}
\caption{Selection of a common detector oversampling factor. 
Complex-object \ac{FRC} values are shown as a function of the oversampling factor \(S\) for the 1-, 2-, and 4-beam configurations. Markers and error bars show the reconstruction results and their uncertainties, and orange curves show quadratic fits. Dashed horizontal lines indicate the fitted maximum \ac{FRC}, while dotted lines indicate one standard deviation below the maximum. Shaded regions show the corresponding \(1\sigma\) intervals in \(S\). The fitted optimal values were 3.67, 3.13, and 3.59 for the 1-, 2-, and 4-beam configurations, respectively. The selected value \(S=3.28\) lies within all three intervals.}
\label{fig:oversampling_scan}
\end{figure}

We evaluated reconstruction quality at different oversampling factors for the 1-, 2-, and 4-beam configurations.
A fixed scan step was used within each configuration, although different scan steps were tested for different numbers of beams. 
For each configuration, the \ac{FRC} of the complex object was evaluated as a function of \(S\), and the resulting values were fitted with a quadratic function,
\begin{equation}
f(S)=aS^2+bS+c.
\label{eq:oversampling_fit}
\end{equation}
The fitted optimum was calculated from the vertex,
\begin{equation}
S_{\mathrm{opt}}=-\frac{b}{2a}.
\label{eq:optimal_oversampling}
\end{equation}
The uncertainty in the maximum \ac{FRC} was propagated through the fitted model. 
The \(1\sigma\) interval in \(S\) was defined by the two intersections between the fitted curve and \(f(S_{\mathrm{opt}})-\sigma_{\mathrm{FRC}}\).
The fitted $1\sigma$ intervals overlapped over $S=3.17$--$3.79$, indicating a common sampling regime for all tested configurations (Fig.~\ref{fig:oversampling_scan}). 
We therefore selected $S=3.28$ for all subsequent simulations.

The configuration used for all subsequent analyses had a detector pixel size of \SI{75}{\micro\metre}, a wavelength of \SI{9.53e-11}{\metre} (approximately \SI{13}{\kilo\electronvolt}), and an object-to-detector distance of \SI{6.25}{\metre}. Both the diffraction patterns and probe arrays contained $200\times200$ pixels, yielding an object-plane pixel size of \SI{3.97e-8}{\metre}.

\subsection{Reconstruction and evaluation}

All datasets were reconstructed using PtyPy~\cite{enders2016computational}, with 1000 initial iterations using the difference map (DM) algorithm~\cite{thibault2009probe} followed by 3000 iterations of the maximum likelihood (ML) algorithm~\cite{thibault2012maximum}.
The simulated probes provided the initial probe estimates and were left free to update throughout all iterations of the reconstruction.
Each object region was initialized with unit amplitude and independently sampled random phases  uniformly distributed between $-\pi/2$ and $\pi/2$ radians.
Identical reconstruction parameters were used for all datasets.

We evaluated reconstruction quality using \ac{FRC} between each reconstructed complex object region and its corresponding ground truth~\cite{van2005fourier,banterle2013fourier}.
The achieved resolution was defined by the highest spatial frequency at which the \ac{FRC} remained above the one-bit threshold~\cite{van2005fourier,van2017reassessing}.
For \ac{MBP}, these resolution frequencies were averaged across the reconstructed object regions to obtain a single value for the whole \ac{MBP} reconstruction.
A reconstruction was considered successful when this value reached at least 0.67 times the Nyquist frequency~\cite{van2005fourier}.
For each number of beams, we identified the largest tested scan step size that still satisfied this criterion, thereby determining the minimum required number of diffraction patterns.
We denote this number by $M_{\mathrm{SBP}}$ for a single object region imaged using \ac{SBP}, and by $M_{\mathrm{MBP}}$ for the regions imaged simultaneously using \ac{MBP}.

\section{Results}

We first compare the diffraction pattern requirements of \ac{MBP} and \ac{SBP} at the same target resolution of 0.67 times the Nyquist frequency to quantify the acquisition gain.
We then examine how this gain changes with the chosen detector cropping and with the relative target resolution.

\subsection{Resolution-matched gain}

\begin{figure}[ht!]
\centering\includegraphics[width=\linewidth]{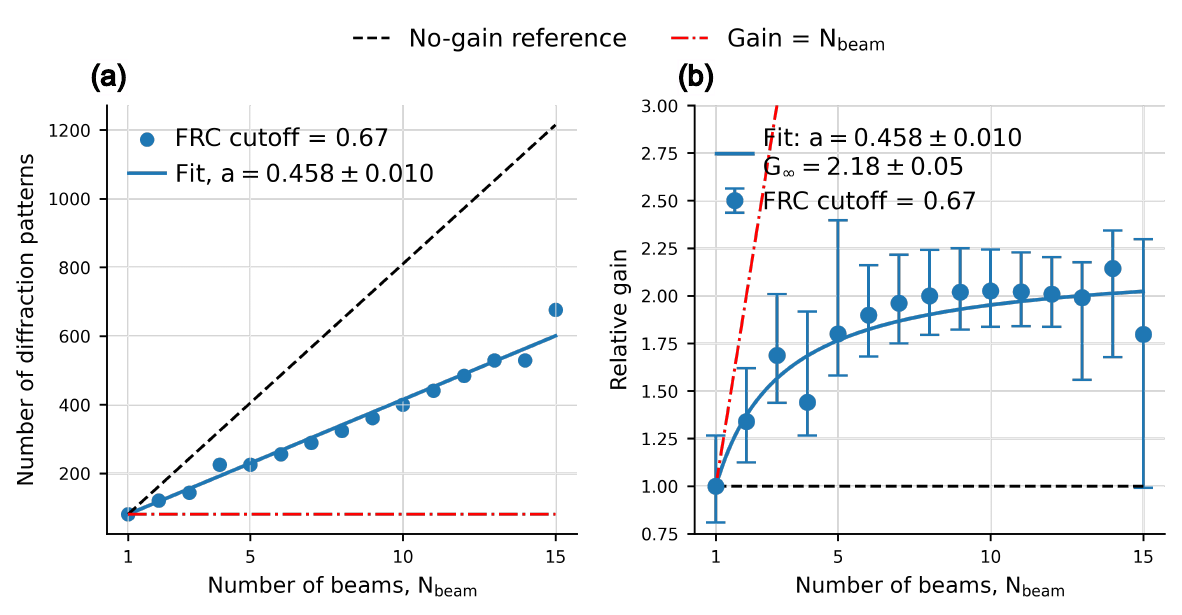}
\caption{Resolution-matched diffraction pattern requirements and relative gains for fully overlapped \ac{MBP}.
(a) Required diffraction pattern number as a function of the number of beams.
(b) Corresponding gain relative to separate \ac{SBP} scans of the same object regions at the same resolution.
The curve in (b) was calculated using Eq.~\ref{eq:relative_gain} using the value of \(a\) obtained from the fit in (a), with no additional fitting to the gain data.
Error bars indicate the uncertainty arising from the discrete sampling of scan step size.
Across all panels, the black dashed no-gain reference (\(G=1\)) indicates that \ac{MBP} requires the same number of diffraction patterns as \(N_{\mathrm{beam}}\) repeated \ac{SBP} scans. 
The red dash-dotted \(G=N_{\mathrm{beam}}\) reference indicates that \ac{MBP} requires the same number of diffraction patterns as a single \ac{SBP} scan. 
}
\label{fig:mbp_tendency}
\end{figure}

At the target resolution of 0.67 times the Nyquist frequency, the required pattern number $M_{\mathrm{MBP}}$ increased approximately linearly with the number of beams as shown in Fig.~\ref{fig:mbp_tendency}(a).
We described this dependence by the following empirical linear model:
\begin{equation}
M_{\mathrm{MBP}}(N_{\mathrm{beam}})
=
M_{\mathrm{SBP}}
\left[
1+a\left(N_{\mathrm{beam}}-1\right)
\right],
\label{eq:mbp_pattern_model}
\end{equation}
where \(a\) is the normalized increase in the required number of \ac{MBP} diffraction patterns per added beam.

To quantify the acquisition gain, we compared \ac{MBP} with $N_{\mathrm{beam}}$ separate \ac{SBP} scans covering the same total object area at the same resolution.
These reference scans require a total of $N_{\mathrm{beam}}M_{\mathrm{SBP}}$ diffraction patterns, giving
\begin{equation}
G
=
\frac{N_{\mathrm{beam}}M_{\mathrm{SBP}}}
{M_{\mathrm{MBP}}}
=
\frac{N_{\mathrm{beam}}}
{1+a\left(N_{\mathrm{beam}}-1\right)}~,
\label{eq:relative_gain}
\end{equation}

Here, \(G=1\) corresponds to \(M_{\mathrm{MBP}}=N_{\mathrm{beam}}M_{\mathrm{SBP}}\), indicating that \ac{MBP} requires the same total number of diffraction patterns as the separate \ac{SBP} scans. 
Values of \(G>1\) indicate a reduction in the required pattern count by a factor of \(G\) for the same total object area and target resolution. 
In particular, \(G=N_{\mathrm{beam}}\) corresponds to \(M_{\mathrm{MBP}}=M_{\mathrm{SBP}}\), representing ideal linear scaling of the gain with the number of beams. 
In this case, \ac{MBP} images all \(N_{\mathrm{beam}}\) regions using the same number of diffraction patterns as a single \ac{SBP} reference scan.
The dependence of the gain on the numebr of beams is governed by the fitted parameter \(a\).
For \(0<a<1\), the gain increases monotonically with \(N_{\mathrm{beam}}\) and asymptotically approaches \(G_{\infty}=1/a\). 
The gain values calculated directly from the simulations were consistent with the curve predicted by Eq.~\ref{eq:relative_gain} using the value of \(a\) obtained from the linear fit of \(M_{\mathrm{MBP}}\), as shown in Fig.~\ref{fig:mbp_tendency}(b).

The fitted coefficient was $a=0.458\pm0.010$, corresponding to a predicted asymptotic gain of $G_{\infty}=2.18\pm0.05$.
Thus, the fitted model predicts that, at large numbers of beams, fully overlapped \ac{MBP} requires approximately half the estimated \ac{SBP} pattern number for the same total area and resolution.

\subsection{Gain dependence}

The gain reported above was obtained using the full diffraction patterns and an \ac{FRC} cutoff of 0.67 times the Nyquist frequency.
We next examined whether this gain changed when the recorded reciprocal-space range or the resolution criterion was changed.

We first reduced the recorded reciprocal-space range by centrally cropping the simulated diffraction patterns from $200\times200$ to $100\times100$ pixels.
No additional photon normalization, noise generation, or intensity correction was applied after cropping.
The simulated probe and ground-truth object were resampled to match the real-space sampling of the cropped data.
Cropping reduced the maximum recorded spatial frequency, \(q_{\max}\), and doubled the real-space pixel size while preserving the physical field of view.
The \ac{FRC} cutoff of 0.67 times the Nyquist frequency therefore corresponded to the same normalized resolution criterion, but not the same absolute spatial resolution.
For the cropped data, \ac{MBP} and \ac{SBP} were evaluated using the same criterion.

\begin{figure}[!htbp]
\centering
\centering\includegraphics[width=\linewidth]{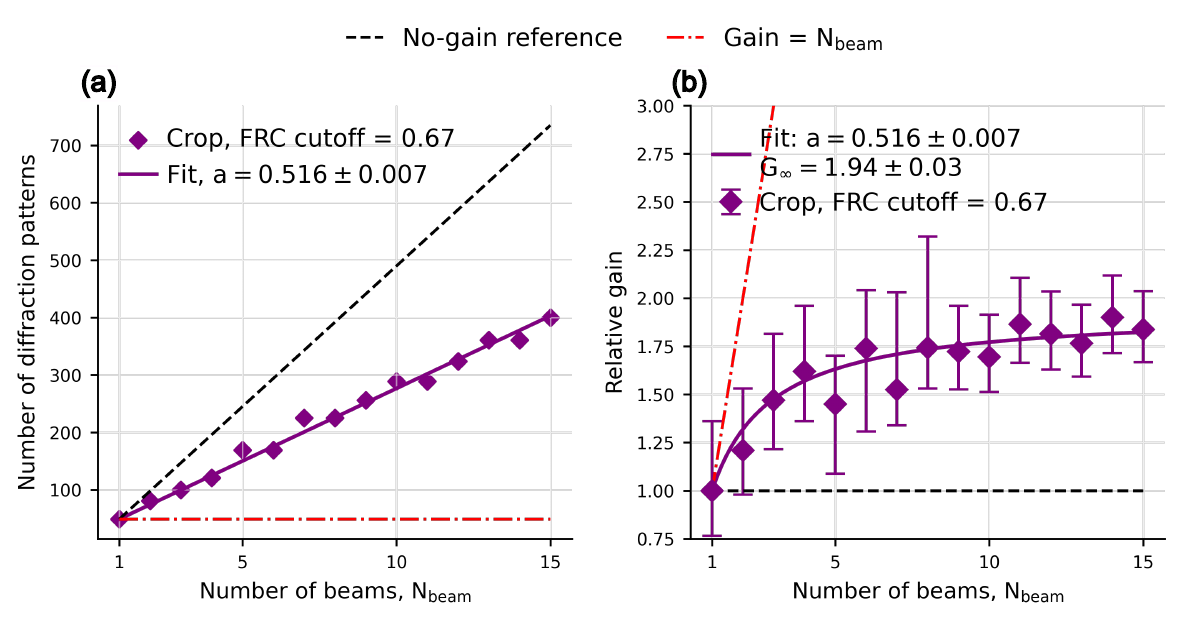}
\caption{Sensitivity of the \ac{MBP} gain to detector $q$ range.
(a) Required diffraction pattern number as a function of the number of beams after cropping the diffraction patterns to $100\times100$ pixels.
(b) Corresponding gain.
Markers, fits, error bars, and reference lines follow the same definitions as in Fig.~\ref{fig:mbp_tendency}.}
\label{fig:mbp_tendency_crop}
\end{figure}

After cropping, $M_{\mathrm{MBP}}$ remained approximately linear in $N_{\mathrm{beam}}$ and was well described by Eq.~\ref{eq:mbp_pattern_model} [Fig.~\ref{fig:mbp_tendency_crop}(a)].
The fitted value of $a$ increased from $0.458\pm0.010$ for the full diffraction patterns to $0.516\pm0.007$ after cropping.
The predicted asymptotic gain consequently decreased from $2.18\pm0.05$ to $1.94\pm0.03$, consistent with the gains calculated from the cropped data [Fig.~\ref{fig:mbp_tendency_crop}(b)].
These results show that the observed gain depends on the retained detector range.

\begin{figure}[ht!]
\centering\includegraphics[width=\linewidth]{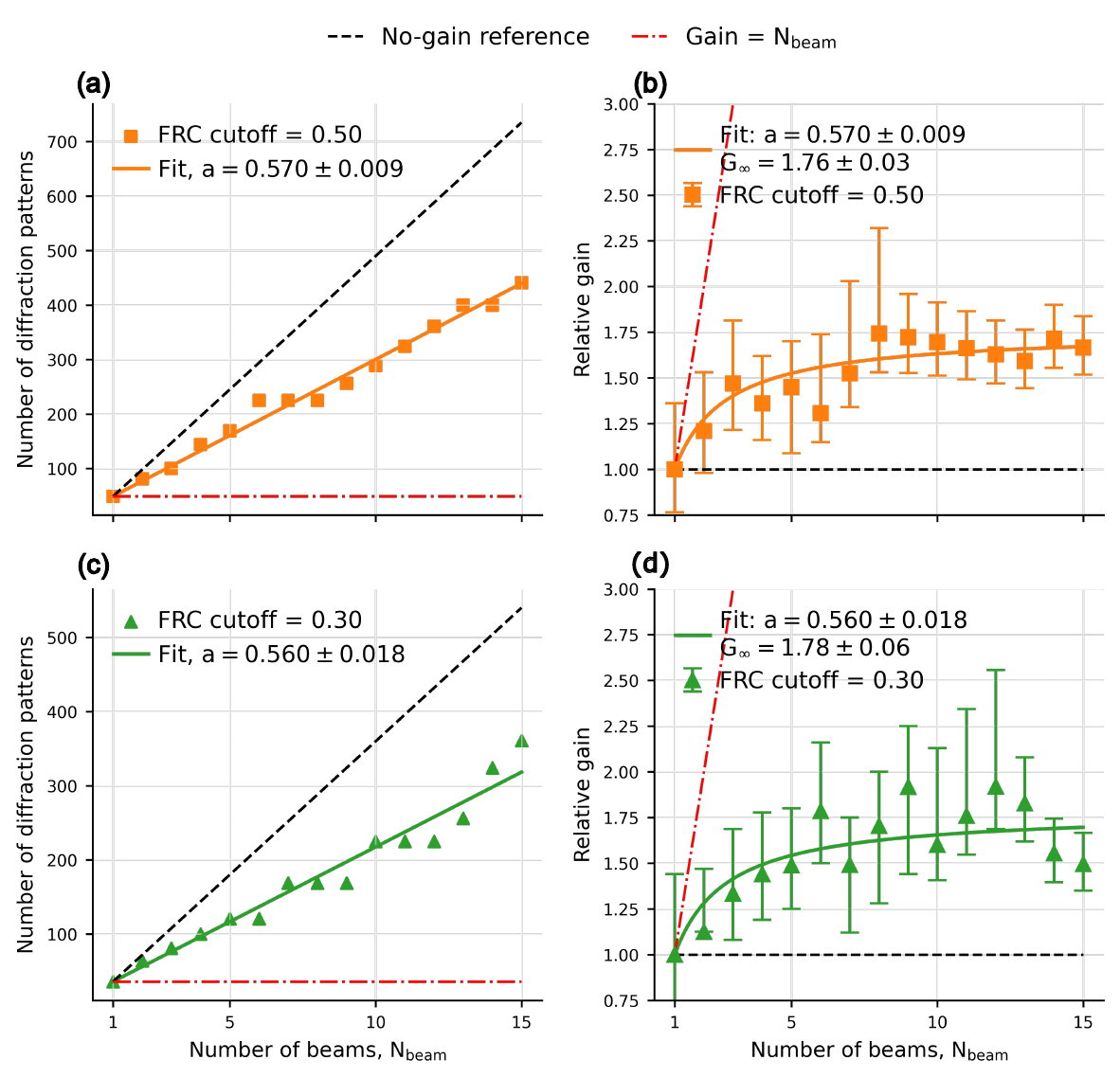}
\caption{Sensitivity of the \ac{MBP} gain to the resolution criterion. 
(a,b) Required diffraction pattern number and corresponding gain at an \ac{FRC} cutoff of 0.50 times the Nyquist frequency. 
(c,d) Corresponding results at a cutoff of 0.30 times the Nyquist frequency.
Markers, fits, error bars, and reference lines follow the same definitions as in Fig.~\ref{fig:mbp_tendency}.}
\label{fig:mbp_dif}
\end{figure}

Because cropping also changed the absolute resolution corresponding to the \ac{FRC} cutoff, we next varied the cutoff while retaining the full detector $q$ range.
We repeated the analysis using the full recorded reciprocal-space range with lower \ac{FRC} cutoffs of 0.50 and 0.30 times the Nyquist frequency.
For both cutoffs, $M_{\mathrm{MBP}}$ remained approximately linear in $N_{\mathrm{beam}}$ [Fig.~\ref{fig:mbp_dif}(a,c)].
Both yielded larger fitted values of $a$ and smaller gains than those obtained at the \ac{FRC} cutoff of 0.67 [Fig.~\ref{fig:mbp_dif}(b,d)].
The gains calculated from the simulations followed the curves obtained from Eq.~\ref{eq:relative_gain} using the corresponding fitted values of $a$.
Thus, with the full detector $q$ range retained, the gain also depended on the chosen resolution criterion.

Together, these results show that the resolution-matched gain depends on the recorded reciprocal-space range and the chosen resolution criterion, rather than on the number of beams alone.

\section{Discussion}

Our results show that fully overlapped \ac{MBP} can reduce the diffraction pattern number relative to the estimated \ac{SBP} requirement for the same total area and resolution.
However, the fitted coefficient $a$ and the corresponding gain changed with the choices of the detector cropping and the \ac{FRC} cutoff.
The information oversampling ratio in Eq.~\ref{eq:info_ratio} counts the pixels of the recorded diffraction pattern relative to unknown object and probe variables.
It does not account for how the photon intensity is distributed within the recorded diffraction pattern or which spatial frequencies must be reconstructed.
These differences motivate examining the diffraction signal in more detail.

\begin{figure}[!htb]
\centering
\includegraphics[width=\linewidth]{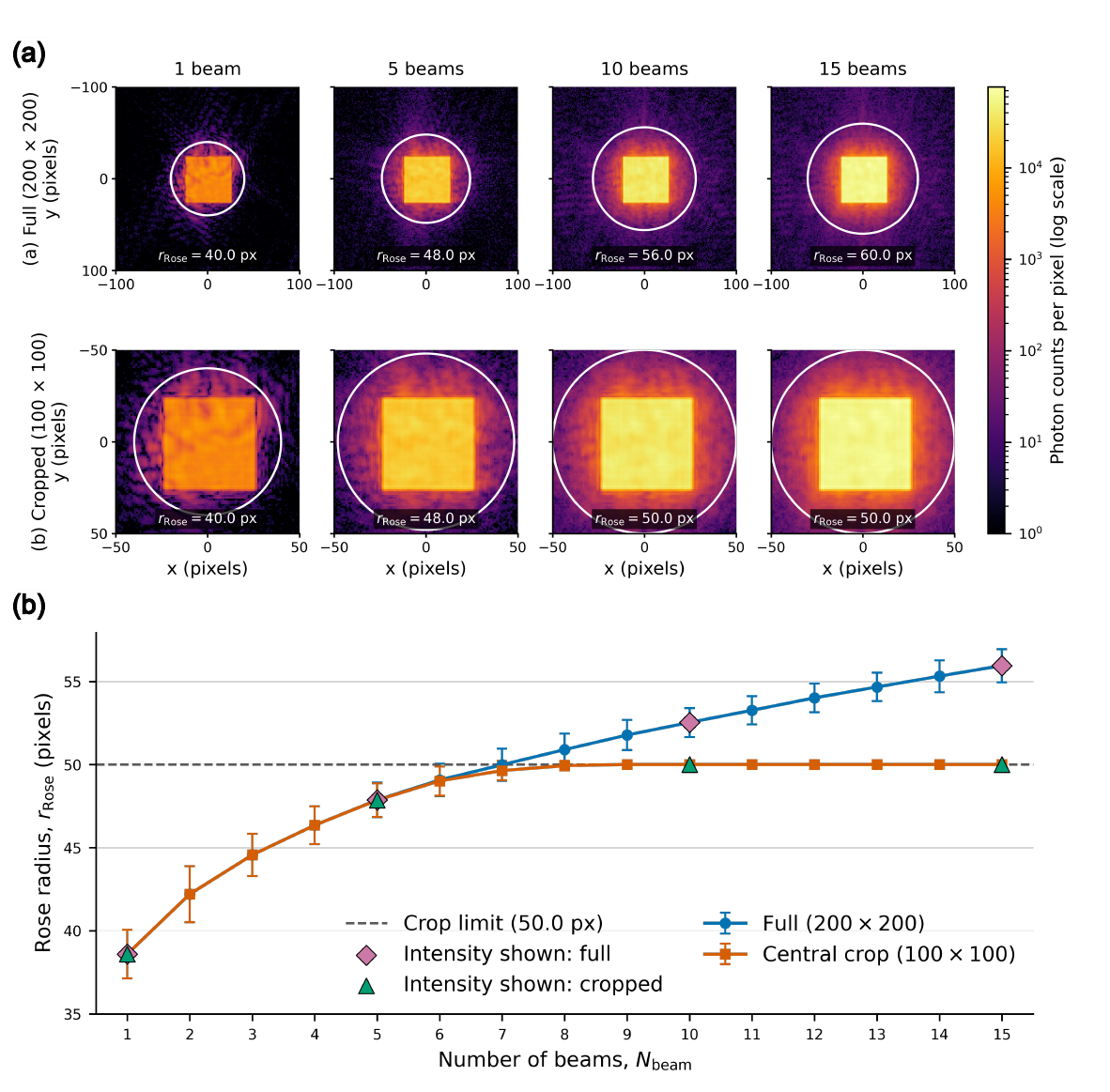}
\caption{Representative diffraction patterns and Rose radii for fully overlapped \ac{MBP}.
(a) Full $200\times200$ diffraction patterns (top row) and central $100\times100$ crops (bottom row) for 1, 5, 10, and 15 beams at the first scan position.
Intensities are displayed on a common logarithmic color scale, and white circles mark the corresponding Rose radii.
Coordinates are detector pixels relative to the array center.
(b) Mean Rose radius as a function of the number of beams, with error bars indicating the standard deviation over all scan positions.
Purple diamonds and green triangles highlight the number of beams shown in (a) for the full and cropped data, respectively.
The dashed line marks the maximum complete radius of the cropped array, $r_{\max}=50$ pixels.}
\label{fig:rose_trend}
\end{figure}

We therefore examined how the detectable diffraction signal changes with the number of beams using the Rose criterion~\cite{burgess1999rose,hsieh2022minimum}.
For the simulated Poisson noise and zero background, we applied a threshold of 25 photons per pixel to the mean photon count in concentric rings.
This threshold corresponds to $\mathrm{SNR}\geq5$ for the photon count per pixel.
The Rose radius was defined as the outermost radius up to which this criterion was continuously satisfied.
The analysis was restricted to the largest complete circle within the detector area.
Representative diffraction patterns illustrate the spatial extent of the detectable signal and the corresponding Rose radii for the full and cropped data (Fig.~\ref{fig:rose_trend}(a)). 
For the full diffraction patterns, the mean Rose radius increased with the number of beams, indicating detectable signal at larger radial distances from the detector center (Fig.~\ref{fig:rose_trend}(b)).
The cropped data followed the same trend until the Rose radius reached the crop boundary, which imposed a plateau rather than indicating saturation of the underlying diffraction signal.
Because the photon number per probe per exposure was fixed, the increase in Rose radius occurred alongside an increase in the total photon number.

These observations suggest that the number of recorded pixels on the detector alone does not fully describe the diffraction information available for reconstruction.
To account for this distinction, we consider an effective diffraction pattern size,
$N_{\mathrm{diff}}^{\mathrm{eff}}(q_{\max})$, within the retained reciprocal-space range:
\begin{equation}
\sigma_{\mathrm{MBP}}^{\mathrm{eff}}
\propto
\frac{
M N_{\mathrm{diff}}^{\mathrm{eff}}(q_{\max})
}{
N_{\mathrm{beam}}N_{\mathrm{unknown,loc}}
}.
\label{eq:info_ratio_redefined}
\end{equation}
The Rose radius characterizes the detectable radial extent, but does not directly measure $N_{\mathrm{diff}}^{\mathrm{eff}}(q_{\max})$ or the number of independent reconstruction constraints.
Its increase is consistent with more of the recorded reciprocal-space range containing detectable signal at large numbers of beams.
Cropping removes part of this signal, whereas a lower \ac{FRC} cutoff reduces the spatial frequency range required to reach the resolution target.

These observations define the experimental regime in which fully overlapped \ac{MBP} is most beneficial. 
Its advantage is expected to be largest when the photon counts from a single beam are insufficient over the \(q\) range required to reach the target resolution. 
In this regime, incoherent summation increases the high-\(|\mathbf{q}|\) signal above the detection threshold and allows these measurements to contribute additional constraints. 
We refer to this effect as the effective utilization of photon information. 
Fully overlapped \ac{MBP} is therefore particularly useful when the target resolution is limited by photon statistics over the required \(q\) range.
This effective utilization of photon information does not imply improved dose efficiency. 
Although \ac{MBP} with \(G>1\) required fewer diffraction patterns than \(N_{\mathrm{beam}}\) repeated \ac{SBP} scans, each \ac{MBP} beam had the same photon count per exposure as the beam used in \ac{SBP}.
Consequently, each object region received \(N_{\mathrm{beam}}/G\) times the exposure of a resolution-matched \ac{SBP} scan.
The gain therefore reflects improved acquisition efficiency rather than more information obtained per photon.
This trade-off may be advantageous when acquisition speed or temporal resolution is limiting, but less favorable for specimens that are sensitive to the increased radiation dose. 
This interpretation clarifies the limitations of the fully overlapped \ac{MBP} geometry.

The conservative nature of our benchmark becomes clear when considering the additional reconstruction challenge introduced by multiplexing~\cite{batey2014information,bevis2018multiple}.
Compared with \ac{SBP}, \ac{MBP} requires sufficient information redundancy not only to constrain the multiple object and probe unknowns, but also to disentangle the diffraction intensity contributions from different beams.
Full~\cite{roper2025stereo} or partial~\cite{penagos2025multiplexing,karl2015spatial} separation of the diffraction patterns on the detector provides additional spatial information for distinguishing the diffraction intensity contributions from different beams.
This additional information can reduce the scan redundancy requirements to reach a given reconstruction resolution, corresponding to a smaller \(a\) and a larger gain than achieved in the fully overlapped geometry.
To describe this effect, we introduce a factor \(\beta_{\mathrm{det}}\) to quantify the degree of overlap among the individual diffraction patterns.
Let \(A_{\beta_i}\) denote the detector area over which exactly \(i\) beam diffraction patterns overlap.
Here \(i\) is the number of beam diffraction patterns contributing to that detector region, and \(A_{\mathrm{det}}=\sum_{i=1}^{N_{\mathrm{beam}}}A_{\beta_i}\) is the total recorded detector area.
A detector plane separation factor can then be written as
\begin{equation}
\beta_{\mathrm{det}}
=
\frac{N_{\mathrm{beam}}}{A_{\mathrm{det}}}
\sum_{i=1}^{N_{\mathrm{beam}}}
\frac{A_{\beta_i}}{i}.
\label{eq:detector_separation_gain}
\end{equation}
Combining this detector plane separation factor with the photon distribution effect gives
\begin{equation}
\sigma_{\mathrm{MBP}}^{\mathrm{eff}}
\propto
\beta_{\mathrm{det}}
\frac{
M N_{\mathrm{diff}}^{\mathrm{eff}}(q_{\max})
}{
N_{\mathrm{beam}}N_{\mathrm{unknown,loc}}
}.
\label{eq:effective_info_ratio}
\end{equation}
Equation~\ref{eq:effective_info_ratio} therefore separates two sources of usable information: \(N_{\mathrm{diff}}^{\mathrm{eff}}(q_{\max})\) describes the effective diffraction information retained within the recorded-reciprocal space range, whereas \(\beta_{\mathrm{det}}\) describes the additional spatial information provided by detector plane separation. 
For the fully overlapped \ac{MBP} geometry, \(\beta_{\mathrm{det}}=1\), and the gain arises entirely from effective utilization of photon information within the recorded \(q\) range. 
In the limit of complete separation, \(\beta_{\mathrm{det}}=N_{\mathrm{beam}}\), which fully offsets the \(N_{\mathrm{beam}}\) factor and maximizes the contribution from detector plane separation.
Partial separation gives
\(1<\beta_{\mathrm{det}}<N_{\mathrm{beam}}\),  and the gain reflects both effective utilization of photon information and the additional spatial information provided by detector plane separation. 
This framework explains why partially~\cite{penagos2025multiplexing} or fully separated geometries~\cite{roper2025stereo} can achieve, and have been shown to achieve gains beyond the fully overlapped benchmark.

\section{Conclusion}

In summary, we quantified the information redundancy requirement of fully overlapped \ac{MBP} under a fixed reconstruction resolution target based on \ac{FRC} evaluation.
The required pattern number increased approximately linearly with \(N_{\mathrm{beam}}\), yielding a fitted asymptotic gain of \(G_{\infty}=1/a\).
Here, \(a\) is not a universal constant but a condition-dependent coefficient.
In the present simulations, it depended on the retained detector \(q\) range and the reconstruction evaluation criterion; more generally, it may also depend on photon statistics, probe diversity, detector-plane overlap, and the reconstruction procedure.
Under the main \ac{FRC} cutoff of 0.67 times the Nyquist frequency using the one-bit threshold, fully overlapped \ac{MBP} required approximately half as many diffraction patterns as the total \(N_{\mathrm{beam}}M_{\mathrm{SBP}}\) patterns required for \(N_{\mathrm{beam}}\) separate resolution-matched \ac{SBP} scans.
Detector cropping and lower \ac{FRC} cutoffs produced larger values of \(a\) and smaller gains than those obtained under the main simulation conditions.
This dependence indicates that the gain is governed not by \(N_{\mathrm{beam}}\) alone, but also by the amount of usable diffraction information retained in each measurement.
Because the fully overlapped geometry provides no detector plane separation between beam diffraction patterns, the gain observed here should be interpreted as a conservative benchmark for \ac{MBP} rather than an upper limit.
Partial or complete detector plane separation provides additional information about the origin of the measured diffraction intensities and therefore offers a route toward larger gains with \ac{MBP}.

\section*{Acknowledgements}
This work was supported by the Röntgen-Ångström Cluster (RÅC) grant (VR 2021-05975). 
The authors thank Tang Li and Mikhail Lyubomirskiy for valuable discussions and constructive suggestions.

\section*{Disclosures}

The authors declare no conflicts of interest. 
The simulated raw data and the reconstructed data, including evaluation files, will be made available in a public repository upon publication.


\bibliography{sample}

\end{document}